\documentclass[traditabstract]{aa} % for the abstract without structuration 
\bibpunct{(}{)}{;}{a}{}{,}

\newcommand{\ergps}{erg\thinspace s$^{-1}$}
\newcommand{\phpspsqcm}{ph\thinspace cm$^{-2}$\thinspace s$^{-1}$}
\newcommand{\ergpspsqcm}{erg\thinspace s$^{-1}$\thinspace cm$^{-2}$}
\newcommand{\psqcm}{cm$^{-2}$}
\newcommand{\nH}{$N_{\rm H}$}
\newcommand{\Msun}{$M_{\odot}$}
\newcommand{\Lsun}{$L_{\odot}$}
\newcommand{\cps}{ct\thinspace s$^{-1}$}

\usepackage[varg]{txfonts}
\usepackage{graphicx}
\begin{document}

\title{A Compton-thick nucleus in the dual active galactic nuclei of Mrk~266}

%   \subtitle{}

\author{K. Iwasawa\inst{1,2}
\and
C. Ricci\inst{3,4,5}
\and
G. C. Privon\inst{6,7}
\and
N. Torres-Alb\`a\inst{8}
\and
H. Inami\inst{9}
\and
V. Charmandaris\inst{10,11}
\and
A. S. Evans\inst{6,12}
\and
J.~M.~Mazzarella\inst{13}
\and
T. D\'iaz-Santos\inst{3,14,11}
%\fnmsep\thanks{Just to show the usage
%          of the elements in the author field}
}

\institute{Institut de Ci\`encies del Cosmos (ICCUB), Universitat de Barcelona (IEEC-UB), Mart\'i i Franqu\`es, 1, 08028 Barcelona, Spain
         \and
ICREA, Pg. Llu\'is Companys 23, 08010 Barcelona, Spain
\and
%Department of Physics and Astronomy, University of California, 4129 Frederick Reine%s Hall, Irvine, CA, 92697, USA
%\and
N\'ucleo de Astronomía de la Facultad de Ingeniería y Ciencias, Universidad Diego Portales, Av. Ejército Libertador 441, Santiago, Chile
\and
Kavli Institute for Astronomy and Astrophysics, Peking University, Beijing 100871, China
\and
Department of Physics and Astronomy, George Mason University, MS 3F3, 4400 University Drive, Fairfax, VA 22030, USA
Department of Astronomy, University of Florida, 211 Bryant Space Sciences Center, Gainesville, FL 32611, USA
\and
National Radio Astronomy Observatory, 520 Edgemont Road, Charlottesville, VA 22903, USA
\and
Department of Astronomy, University of Florida, P.O. Box 112055, Gainesville, FL 32611, USA
\and
Department of Physics and Astronomy, Clemson University, Kinard Lab of Physics, Clemson, SC 29634, USA
\and
Hiroshima Astrophysical Science Center, Hiroshima University, 1-3-1 Kagamiyama, Higashi-Hiroshima, Hiroshima 739-8526, Japan
\and
Department of Physics, University of Crete, GR-71003 Heraklion, Greece
\and
Institute of Astrophysics, Foundation for Research and Technology—Hellas, Heraklion, GR-70013, Greece
\and
Department of Astronomy, University of Virginia, P.O. Box 400325, Charlottesville, VA 22904, USA
\and
IPAC, MC 100-22, California Institute of Technology, Pasadena, CA, 91125, USA
\and
Chinese Academy of Sciences South America Center for Astronomy, National Astronomical Observatories, CAS, Beijing 100101, China
}

%   \date{;}

% \abstract{}{}{}{}{} 
% 5 {} token are mandatory
 
\abstract{
We present the results from our analysis of {\it NuSTAR} data of the luminous infrared galaxy Mrk 266, which contains two nuclei, south-western (SW) and north-eastern (NE), which were resolved in previous {\it Chandra} imaging. Combining this with the {\it Chandra} data, we intepret the hard X-ray spectrum obtained from a {\it NuSTAR} observation to result from a steeply rising flux from a Compton-thick active galactic nuclei (AGN) in the SW nucleus which is very faint in the {\it Chandra} band, confirming the previous claim. This hard X-ray component is dominated by reflection, and its intrinsic 2-10 keV luminosity is likely to be $\sim 1\times 10^{43}$ \ergps. Although it is bright in soft X-ray, only a moderately absorbed NE nucleus has a 2-10 keV luminosity of $4\times 10^{41}$ \ergps, placing it in the low-luminosity AGN class. These results have implications for understanding the detectability and duty cycles of emission from dual AGN in heavily obscured mergers.
}

\titlerunning{Dual AGN in Mrk 266}
\authorrunning{K. Iwasawa et al.}
\maketitle
%
%________________________________________________________________

\section{Introduction}

% Fig. 1 -- HST image + Chandra image

\begin{figure}
\centerline{\includegraphics[width=0.4\textwidth,angle=0]{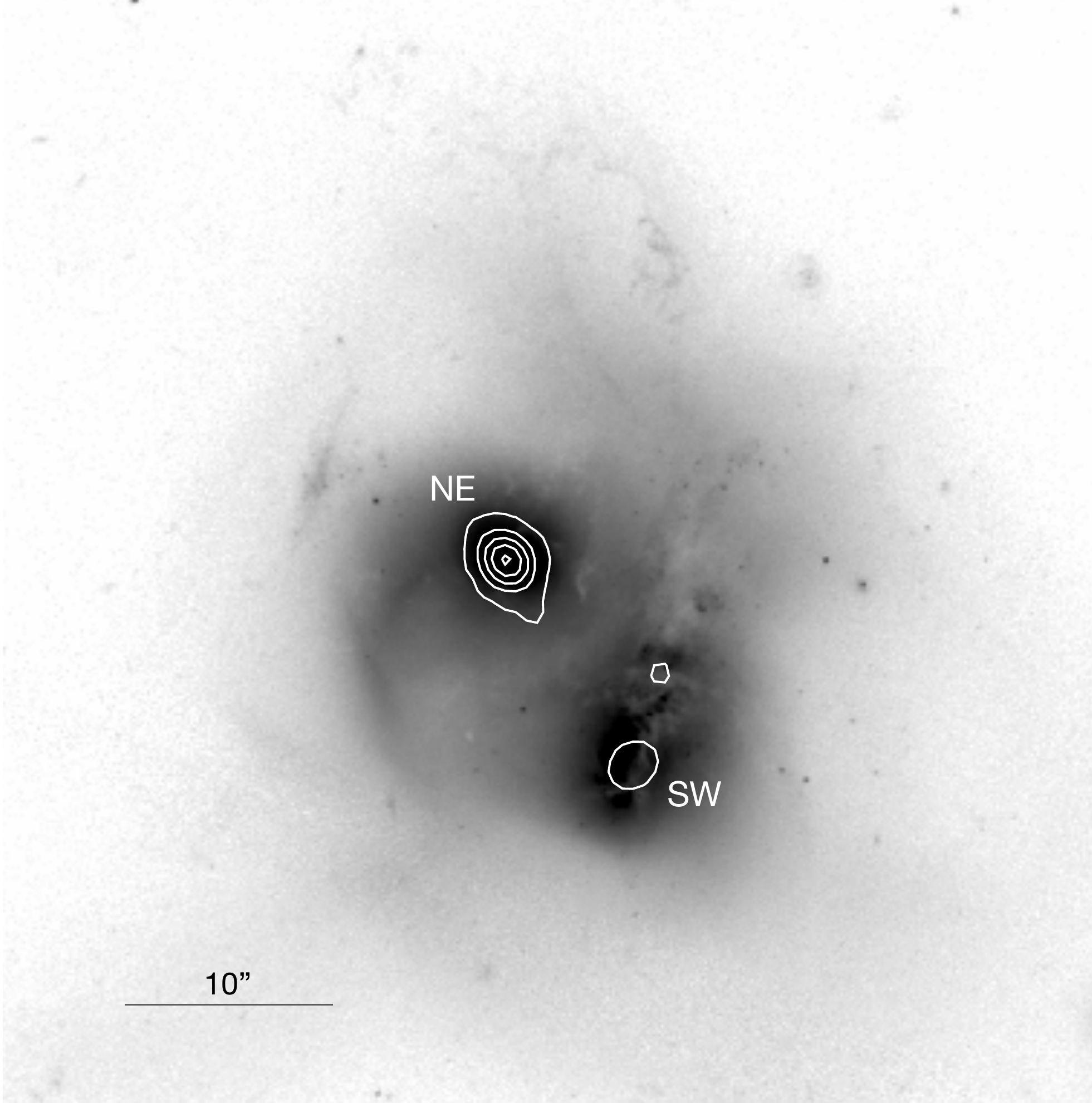}}
\caption{HST ACS/WFC image of Mrk 266 obtained with the F814W filter, overlaid by Chandra 4-7 keV image contours. The five contour levels are linearly spaced between 0.15 to 3.8 counts pixel$^{-1}$. North is up, and east is to the left. The SW and NE nuclei are labelled. The scale bar of 10 arcsec is shown.}
\label{fig:hstimg}
\end{figure}

% Fig. 2 --- All fnspec XMM+Nu, Chandra NE and Sw 
\begin{figure}
\centerline{\includegraphics[width=0.45\textwidth,angle=0]{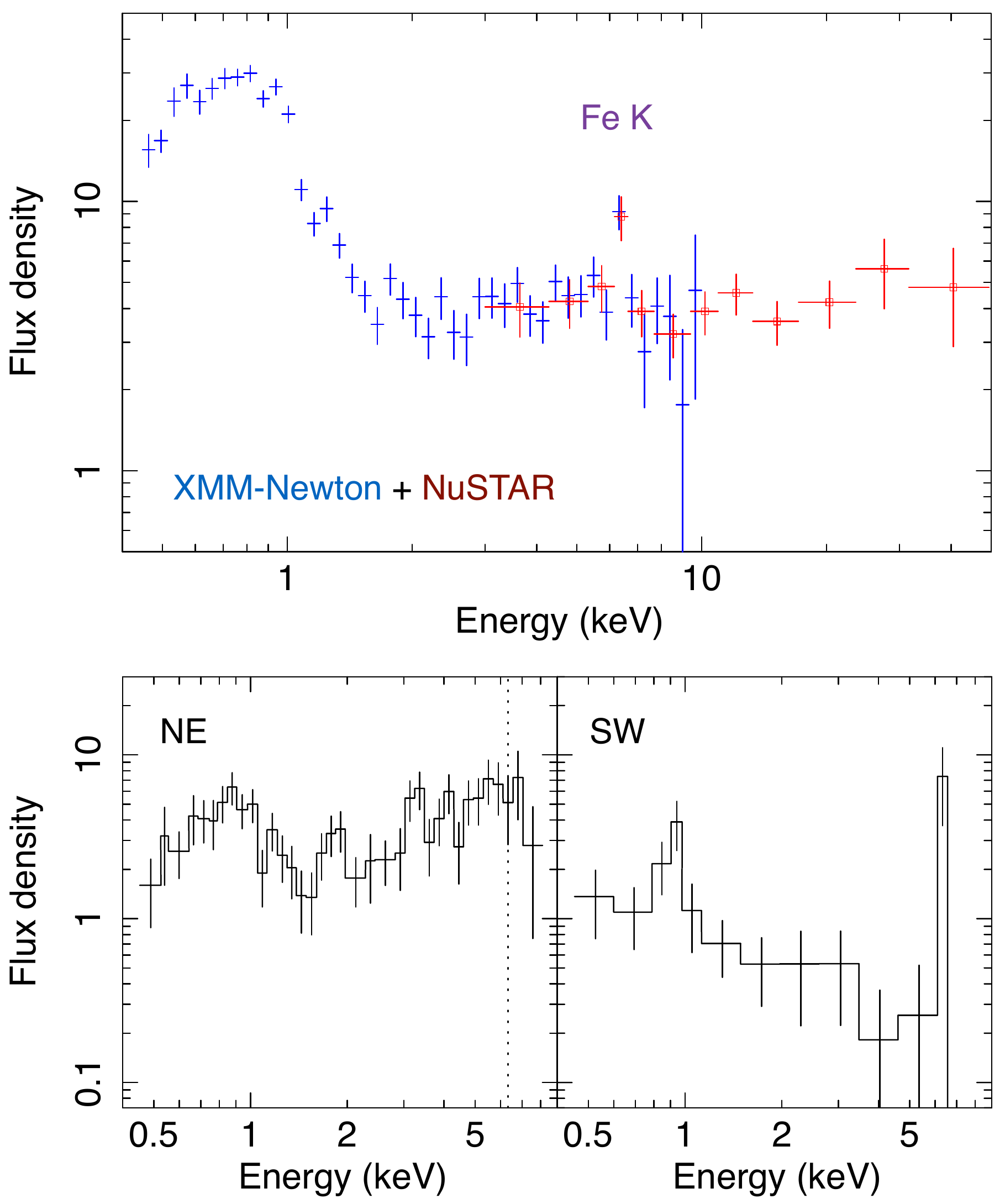}}
\caption{Upper panel: Flux density spectrum of integrated emission of
  Mrk 266, containing both NE and SW nuclei. Data from XMM-Newton
  (blue) and NuSTAR (red squares) are plotted. The flux density is in
  units of $10^{-14}$ erg\thinspace s$^{-1}$\thinspace
  cm$^{-2}$\thinspace keV$^{-1}$. For display purposes, the NuSTAR data used for the spectral analysis were rebinned further. The Fe K emission line feature at 6.5
  keV is marked. Bottom panel: Spectra of the NE and SW nuclei
  obtained from the Chandra ACIS. Since Fe K line emission is absent
  in the NE spectrum (the line energy is denoted by the vertical
  dotted line), the Fe K line seen in the XMM-Newton and NuSTAR
  spectra seems to originate predominantly from the SW.}
\label{fig:multispec}
\end{figure}

Mrk 266 (= NGC~5256) is a major merger of two galaxies with nearly equal masses with a nuclear separation of 10 arcsec, corresponding to a projected physical distance of 6 kpc at $z=0.0279$. It is a luminous infrared galaxy (LIRG) with $L_{{8-1000\mu}{\rm m}} = 3.6\times 10^{11}$\Lsun\ and is a member of the Great Observatories All-sky LIRG Survey \citep[GOALS, ][]{armus09}. While a significant portion of the large IR luminosity originates in intense star formation, both galaxies host active galactic nuclei (AGN). Dual AGN in galaxy mergers are relatively rare in infrared-selected samples such as GOALS \citep[four out of 54 multiple systems,][]{iwasawa11,torres-alba2018,iwasawa18}; this is likely due to heavy obscuration making the dual AGN difficult to detect as opposed to the absence of them. Mrk 266 provides a good opportunity for a detailed study of dual AGN in a luminous IR system. The two AGN, as described in this paper, are, however, found to have rather different characteristics despite their similar host galaxy masses; additionally, heavy obscuration does indeed play a significant role.

Following the extensive multiwavelength analysis of Mrk~266 presented by Mazzarella et al. (2012, hearafter MIV12), here, we give a brief summary of known properties focusing on the two active nuclei residing in the south-western (SW) and north-eastern (NE) galaxies. The galaxies have comparable masses of around (5-6)$\times 10^{10}M_{\odot}$, from which the black hole mass of each nucleus is estimated to be $\sim 2\times 10^8M_{\odot}$, based on their bulge luminosities \citep{marconihunt03}. The individual galaxies have $L_{\rm IR}$ of $2.3\times 10^{11}L_{\odot}$ (SW) and $0.7\times 10^{11} L_{\odot}$ (NE). This seems to be compatible with the Herschel imaging at 70 $\mu $m and 100 $\mu $m, which marginally resolves them \citep{chu2017}. While there is confusion surrounding the reversed classifications given in the literature, the most likely optical classifications of the SW and NE nuclei are Seyfert 2 and LINER, respectively, as supported by works either dedicated to Mrk 266 or small samples including Mrk 266 \citep{osterbrock1983,kollatschny1984,hutchings1988,mazzarellaboroson1993,osterbrockmartel1993,wang1997,wu1998,ishigaki2000}. The mid-IR spectroscopy with the Spitzer IRS, as summarised by \citet{mazzarella12} (references are therein), shows that the spectra of both nuclei are dominated by polycyclic aromatic hydrocarbon (PAH) emission, indicating an intense starburst. However, the presence of AGN is also suggested by high-ionisation lines, such as [Ne {\sc v}]$\lambda 14.3, 24.3\mu $m, and [O {\sc iv}]$\lambda 25.9\mu $m for the SW nucleus and warm dust continuum for the NE nucleus \citep[see also][]{imanishi09}. A Chandra observation, which resolves the two nuclei, shows that a hard-spectrum source is clearly seen at the NE nucleus (Fig. \ref{fig:hstimg}), supporting the presence of a moderately obscured AGN with \nH\ $= (7\pm2)\times 10^{22}$ \psqcm\ \citep{brassington2007,mazzarella12,torres-alba2018}. On the other hand, the SW nucleus is $\sim $13 times fainter than the NE nucleus and barely detected in the Chandra hard-band (4-7 keV) image (Fig. \ref{fig:hstimg}). The SW nucleus is also fainter than the NE nucleus in the optical and ultraviolet (UV), but it becomes progressively brighter at longer wavelengths $\geq 3\mu$m, leading to the SW nucleus to be $\sim $3 times more luminous than the NE nucleus in the entire IR band. The molecular gas mass in the SW nucleus is estimated to be M(H$_2) =3.4\times 10^9 M_{\odot}$, which is $\sim $5 times larger than that of the NE nucleus \citep{imanishi09}. Even though these measurements were obtained from interferometric observations, they are consistent with the single-dish measurement by \citealp{sanders1986}. These characteristics suggest that obscuration might be responsible for the relative faintness of the SW nucleus in X-rays. The XMM-Newton spectrum of Mrk 266, which does not resolve the individual galaxies, shows a strong Fe K line \citep{mazzarella12}. No Fe K line is present in the Chandra spectrum of the NE nucleus, but there is one in the spectrum of the SW nucleus (the Fe K line is, in fact, the major source of 4-7 keV counts detected with Chandra). This led \citet{mazzarella12} to suspect that the SW nucleus contains a Compton-thick AGN, in which the direct X-ray continuum is totally suppressed in the Chandra bandpass and only the Fe K line in its reflected light is observed. It is reminiscent of the double nucleus in another merger system in GOALS, Mrk 273, in which a Compton-thick nucleus is inferred for the nothern nucleus, which is molecular-rich, and the major IR source of the merger system \citep[e.g.][]{u13,iwasawa11,iwasawa18,liu2019}. In this paper, we examine this hypothesis for Mrk 266 using new NuSTAR data which cover the 3-50 keV band, expecting a sharp rise in the X-ray flux of the SW nucleus above 10 keV, which would exceed the flux from the NE nucleus in the NuSTAR band. The cosmology adopted here is $H_0=70$ km s$^{-1}$ Mpc$^{-1}$,
$\Omega_{\Lambda}=0.72$, $\Omega_{\rm M}=0.28$.

\section{Observations}

% Table 1. -- obs log

\begin{table*}
\caption{X-ray observations of Mrk 266.}
\label{tab:obslog}
\centering
\begin{tabular}{lccccc}
  \hline\hline
  Observatory & Date & ObsID & Exposure & Counts & Band \\
  \hline
Chandra & 2001-11-02 & 2044 & 20 & 124 & 3-8 keV \\
XMM-Newton & 2002-05-15 & 0055990501 & 13/18 & 562 & 3-10 keV \\
NuSTAR & 2019-02-08 & 60465005002 & 64 & 441 & 3-50 keV \\
\hline
\end{tabular}
\tablefoot{The 'Exposure' column shows a useful exposure time in each
  observation in units of $10^3$ seconds. The 'Counts' column gives
  background-corrected source counts in the energy band, which is shown in the
  following column. The two exposure times shown for the XMM-Newton
  observation are of the EPIC pn and the two EPIC MOS cameras. The
  source counts for XMM-Newton and NuSTAR are the sums of the three EPIC
  cameras and two FPMs, respectively.}
\end{table*}

Mrk 266 was observed with NuSTAR, XMM-Newton, and the Chandra X-ray
Observatory and the observation log is shown in Table
\ref{tab:obslog}. The NuSTAR data were taken as part of the NuSTAR
Obscured Seyferts Survey (PI: J. Miller) and we retrieved the data
from the public archive. We used the event file processed by the
standard pipeline. Spectral data extracted from a circular aperture with a radius of 30$^{\prime\prime}$ of the two focal plane modules,
FPMA and FPMB, were combined for the analysis presented below. The spectral data of the two modules are in agreement with each other within a statistical error, and the low-energy effective area problem reported for FPMA \citep{madsen2020nustar} seems to have little impact. Since the NE
and SW nuclei are unresolved with the NuSTAR beam, the
spectrum contains emission from both nuclei, as in the XMM-Newton
spectrum. The spectral data obtained from XMM-Newton and Chandra are
the same as those presented in \citet{mazzarella12}.

\section{Results}

The NuSTAR spectrum is in agreement with the XMM-Newton EPIC spectrum in the
overlapping 3-10 keV band, both in continuum and the Fe K line
(Fig. \ref{fig:multispec}). 
The 3-50 keV NuSTAR spectrum can be well described by a power-law of
energy index $\alpha = 0.0\pm 0.1$\footnote{We used energy-index $\alpha $ for a power-law spectral slope since it is appropriate for the flux density spectra shown in this article. It is related to the conventional photon index $\Gamma $ by $\Gamma = 1 + \alpha$.} plus a narrow Gaussian line at $6.5\pm 0.1$
keV for Fe K ($\chi^2 =46.0$ for 51 degrees of freedom). The Fe line intensity is $(2.9\pm 1.0)\times 10^{-6}$ \phpspsqcm, corresponding to EW $= 0.80\pm 0.27$ keV.

A hard spectrum with a slope, as observed at lower energies, would be interpreted as evidence of moderate absorption of \nH $\leq 10^{23}$ \psqcm\  since an intrinsic AGN spectrum is known to have a
well-defined range of slope around $\alpha = 0.8$ with a standard
deviation of $0.15$ \citep[e.g.][]{ueda14,nandrapounds94}.
The measured spectral slope, $\alpha\approx 0.0$, which stretches up to 50
keV, is, however, unusual because a spectral slope would start to approach the intrinsic value towards higher energies as the effect of such moderate absorption diminishes. Therefore,
it is unlikely that a single source, that is the NE nucleus, dominates
the NuSTAR bandpass as it does in the 4-7 keV Chandra image. Instead,
the NuSTAR spectrum strongly suggests, besides the NE source, the
presence of an extra component rising above the Chandra bandpass, which elevates the hard band continuum emission. This is the same line of reasoning used to convey the presence of a Compton-thick AGN in the northern nucleus of Mrk 273 using NuSTAR data \citep{iwasawa18}. As argued in \citet{mazzarella12}, the SW nucleus is a likely source of this extra hard component, which can be attributed to a Compton-thick AGN.

% Fig --- torus 
\begin{figure}
\centerline{\includegraphics[width=0.48\textwidth,angle=0]{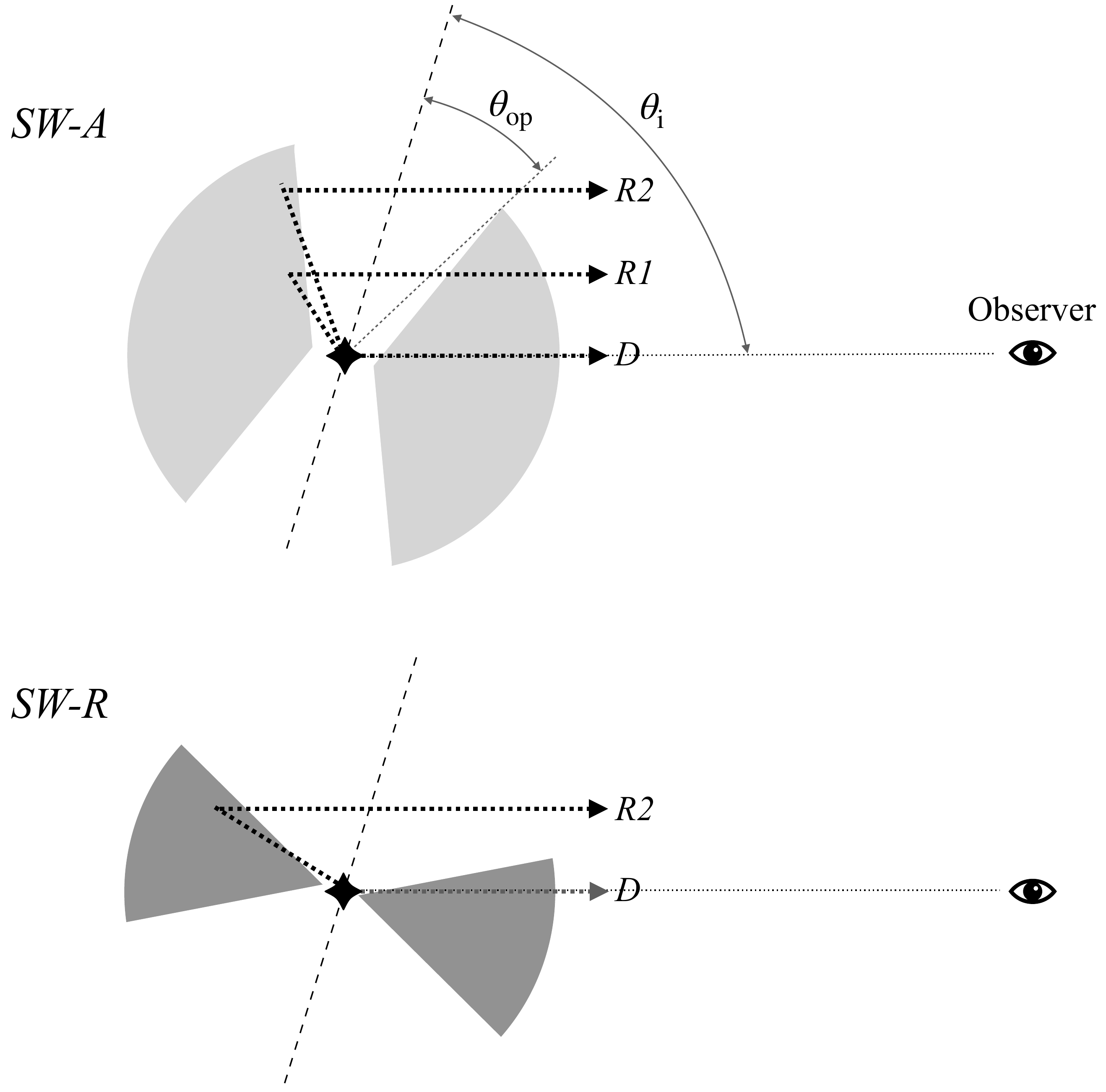}}
\caption{Cross-section view of X-ray absorbing torus configurations assumed for the absroption-dominated ({\it SW-A}) and reflection-dominated ({\it SW-R}) models. The torus parameters of {\tt etorus} given in Table \ref{tab:nu_fit} are illustrated. The inclination angle and half-opening angle are denoted as $\theta_{\rm i}$ and $\theta_{\rm op}$; the labels of $D$, $R1,$ and $R2$ represent direct light from the central X-ray source penetrating through the absorbing column in the line of sight, reflected light coming off from the part of the inner surface of the torus, which is hidden from direct view (`Reflection 1' as defined in \citealp{ikeda09}), and reflected light coming from the visible inner surface (`Reflection 2'), respectively. In {\it SW-A}, an observed spectrum is dominated by $D$ and weaker $R1$, which has practically the same spectral shape as $D$. In {\it SW-R}, $R2$, which has a Compton-scattered low-energy tail, is the dominant component since $D$ (and a minor $R1$ contribution) is strongly suppressed by the large optical depth in the line of sight.}
\label{fig:torus}
\end{figure}

% Table  -- two component fits

\begin{table}
\caption{Two-component fits to the NuSTAR spectrum.}
\label{tab:nu_fit}
\centering
\begin{tabular}{lcc}
  \hline\hline
&  \multicolumn{2}{c}{NE} \\
  \hline
  \nH & \multicolumn{2}{c}{$6.8\times 10^{22}$} \\
  \hline
  &  \multicolumn{2}{c}{SW} \\
  & {\it SW-A} & {\it SW-R} \\
  \hline
  \nH & $1.7^{+1.2}_{-0.6}\times 10^{24}$ & $7\times 10^{24}$\\
  $\theta_\mathrm{op}$ & $30^{\circ}$ & $70^{\circ}$ \\
  $\theta_\mathrm{i}$ & $72^{\circ}$ & $72^{\circ}$ \\
  \hline
  &  \multicolumn{2}{c}{Fe K line} \\
  \hline
  $E$ & $6.49\pm 0.09$ & $6.49\pm 0.09$ \\
  $I$ & $2.6\pm 1.0$ & $2.5\pm 1.0$ \\
  \hline
  $\chi^2$/dof & 46.8/50 & 45.9/51 \\
  \hline
\end{tabular}
\tablefoot{The 3-50 keV NuSTAR spectrum was fitted by NE and SW  composite models with two modelling options for the SW component. 1) {\it SW-A}: absorption-dominated; and 2) {\it SW-R}: reflection-dominated spectrum. The spectral slope for both components were assumed to be $\alpha=0.9$. The SW spectrum was computed by the torus spectrum model {\tt etorus} \citep{ikeda09}. The torus configuration was determined by an opening angle, $\theta_\mathrm{op}$, and inclination, $\theta_\mathrm{i}$. Apart from the absorbing column for {\it SW-A}, free parameters are normalisations of the respective NE and SW continuum and the centroid energy and intensity of a Gaussian for the Fe K line. The values of \nH, line energy ($E$), and intensity ($I$) are in units of \psqcm, keV, and $10^{-6}$ \phpspsqcm, respectively.}
\end{table}

% Table  -- decomposed flux

\begin{table}
\caption{Decomposed fluxes of the NE and SW nuclei.}
\label{tab:flux_2comp}
\centering
\begin{tabular}{lcc}
  \hline\hline
  Band & NE & SW \\
  \hline
  \multicolumn{3}{c}{\it SW-A}\\
  \hline
3-10 keV & 2.4 & 0.15 \\
10-30 keV & 3.1 & 4.2 \\
\hline
\multicolumn{3}{c}{\it SW-R}\\
\hline
3-10 keV & 2.1 & 0.45 \\
10-30 keV & 2.7 & 4.9 \\
\hline
\end{tabular}
\tablefoot{Fluxes for the NE and SW components in the 3-10 keV and 10-30 keV bands were obtained from the spectral decompositions of the NuSTAR data for the two spectral modelling options for SW, {\it SW-A,} and {\it SW-R} (see text and Table \ref{tab:nu_fit}). The fluxes are in units of $10^{-13}$ \ergpspsqcm. The 3-10 keV flux for the SW nucleus is for the continuum only and it does not include the Fe K line flux, which is $0.24\times 10^{-13}$ \ergpspsqcm.}
\end{table}

% two component fits with two SW options
\begin{figure*}
\centerline{\includegraphics[width=0.65\textwidth,angle=0]{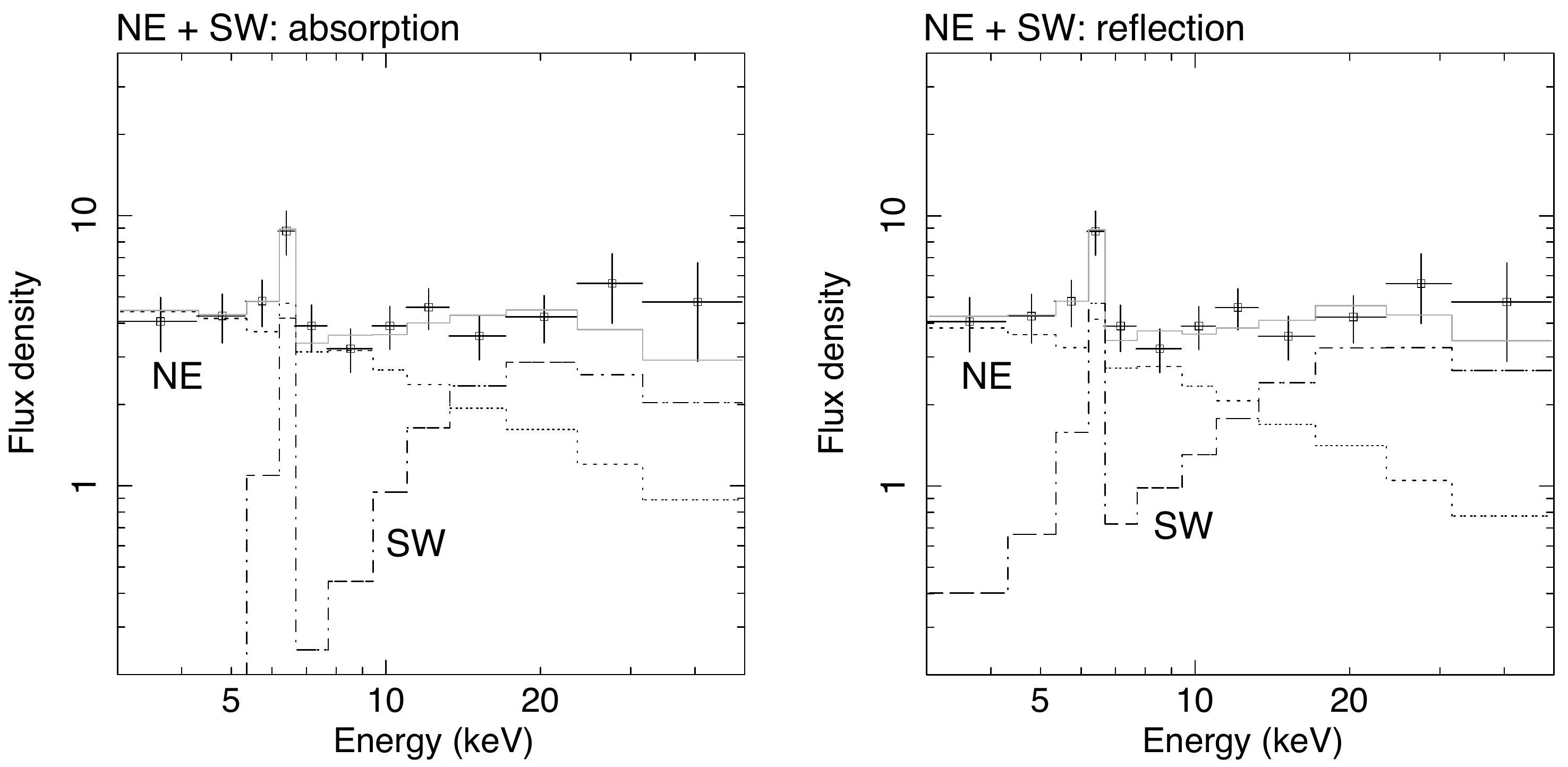}}
\caption{NuSTAR spectrum of Mrk 266 and the best-fitting
  two-component model (grey line). Left: {\it SW-A}; Right: {\it SW-R}. The contributions of the NE (dotted
  line) and SW (dash-dotted line) nuclei are overplotted. Same as in Fig. 2, the NuSTAR data used for the analysis were rebinned further for display purposes. Relative
  contributions of the NE and SW nuclei change dramatically across the energy
  range.}
\label{fig:fn_fit}
\end{figure*}

According to the above hypothesis, we modelled the NuSTAR spectrum with two components from the NE and SW nuclei. To avoid over-fitting, we tested the two-component model, leaving a minimum set of parameters free. Many model parameters were set to reasonable or known values. A first assumption is that both X-ray sources in the NE and SW nuclei have the identical, standard AGN slope of $\alpha = 0.9$. This is the intrinsic slope of the power-law source, not the apparent slope of $\alpha = 0.8$ mentioned above, which is affected by a contribution of reflection from a surrounding medium, as is found in \citet{nandrapounds94}.

The NE spectrum is mildly absorbed. The absorbing column density is found to be \nH\ $= (6.8\pm 3.8)\times 10^{22}$ \psqcm\  by fitting the Chandra NE spectrum and we adopted the best-fit \nH. This leaves the normalisation of the power-law as the only free parameter for the NE component.

We are primarily interested in the SW component. However, as shown below, various configurations of a Compton-thick torus can describe the NuSTAR spectrum, as long as \nH\ $\geq 2\times 10^{24}$ \psqcm\ is met. Therefore we tested the following two modelling options for a Compton-thick AGN: {\it SW-A}, which is a strongly absorbed spectrum, and {\it SW-R}, which is a reflection-dominated spectrum. In both cases, we used {\tt etorus} \citep{ikeda09,awaki09}, which is one of the Monte-Carlo codes, to compute X-ray spectra emerging from an X-ray source surrounded by an absorber with a torus geometry, including Compton-scattered light. The {\tt etorus} offers flexible parameterisations of torus geometry which are similar to those in another recent code {\tt borus02} \citep{balokovic2018}; additionally, \citet{balokovic2018} report that the two codes produce well-matched reflection spectra, apart from fluorescent emission-lines, which are not included in the current version of {\tt etorus}. We modelled the most prominent Fe K line with a narrow Gaussian. The torus parameters were chosen so that a resulting model gives a representative spectral shape for each option. They are illustrated in Fig. \ref{fig:torus} (see Figures 3 and 4 from \citealp{ikeda09} for a visualisation of the simulated spectral components). Since the line-of-sight is obscured, the absorbing torus is expected to be highly inclined and we assume its inclination angle to be $\theta_\mathrm{i} = 72^{\circ}$, which is approximately in the middle of the $45^{\circ}$-$90^{\circ}$ range. In {\it SW-A}, a relatively narrow torus-opening angle of $\theta_\mathrm{op} = 30^{\circ}$ is assumed so that the inner surface of the torus is hardly visible and thus the observed spectrum is dominated by an absorbed component of the central source. The absorbing column and the power-law normalisation were fitted. The absorbing column density is found to be \nH\ $= 1.7^{+1.2}_{-0.6}\times 10^{24}$ \psqcm. In {\it SW-R}, a wide opening angle of $\theta_\mathrm{op} = 70^{\circ}$ and a large \nH\ $= 7\times 10^{24}$ \psqcm\ were assumed. This results in a spectrum being reflection dominated since the highly Compton-thick opacity assures that the direct light from the central source is strongly suppressed and that it has a negligible contribution to the observed spectrum. Whereas the wide $\theta_\mathrm{op}$ makes the inner wall of the torus well exposed to our view; however, the central source is still hidden from direct view by a small margin of $\Delta\theta=2^{\circ}$. The only free parameter is the power-law normalisation.

These spectral modellings are good matches for the NuSTAR data with a comparable quality, as is shown in Fig. \ref{fig:fn_fit} and Table \ref{tab:nu_fit}. The Fe K line EW is found to be $\simeq 0.6\pm 0.23$ keV with respect to the total continuum. We note that depending on the SW modelling options, decomposed fluxes of the NE and SW nuclei vary (Table \ref{tab:flux_2comp}).

The NuSTAR data cannot distinguish between {\it SW-A} and {\it SW-R}. The primary reason is that, although the spectral models for the SW nucleus that were expected from {\it SW-A} and {\it SW-R} differ at some degree at energies below the Fe K band (see Fig. \ref{fig:fn_fit}), this difference is compensated for by varying normalisation of the NE model. Here we investigate if the Chandra data of the SW nucleus can tell the difference between {\it SW-A} and {\it SW-R}, albeit with low source counts. The Chandra data for the SW nucleus have only six counts in 3-7 keV and there are no counts at higher energies; the six counts are composed of four counts for the Fe K line and two counts in the 4-5 keV continuum. The {\it SW-A} model has a sharp decline below 6 keV and its expected count rate in 4-5 keV is $5\times 10^{-6}$ \cps\ while the {\it SW-R} model has a Compton-scattered tail towards lower energies giving an expected count rate of $5\times 10^{-5}$ \cps. The observed count rate $1.0\times 10^{-4}$ \cps lies closer to the latter. Among 1000 simulations of Chandra data based on the absorption model, there were no incidences of two or more counts in 4-5 keV, which, in contrast, occur for 19\% of simulations under the reflection model. This seems to favour the refelction model unless the 4-5 keV counts come from any other external sources, for example, X-ray binary emission from a starburst, which cannot be ruled out.

%The EW of the Fe K line with respect to the SW continuum is 2.2 (1.3-3.1) keV.

Estimating an intrinsic 2-10 keV luminosity, $L_{\rm X}^{\prime}$, of an absorbed X-ray source is straightforward when its absorption opacity is significantly smaller than a Thomson depth or \nH\ $\ll 10^{24}$ \psqcm. For example, in the case of the NE X-ray source, $L_{\rm X}^{\prime}{\rm (NE)}$ is (3.8-4.4)$\times 10^{41}$ \ergps, depending on the choice of the spectral model for the SW nucleus. In the Compton-thick regime, the absorption correction has a dependency on an assumed torus geometry \citep{matt1999}. This applies to the absorption-dominated model ({\it SW-A}) for the SW nucleus. In our assumed torus geometry, which has a large covering fraction (Table \ref{tab:nu_fit}), $L_{\rm X}^{\prime}$(SW) $=2.0\times 10^{42}$ \ergps. However, a similar absorption-dominated spectrum can be obtained with a smaller covering torus with, for instance, $\theta_{\rm op} = 70^{\circ}$ if the torus is precisely edge-on ($\theta_{\rm i}\simeq 90^{\circ}$). In this configuration, since less light is Compton-scattered into the line of sight from a small-covering torus than from that of large-covering, the (absorbed) illuminating source has to be more luminous, resulting in a factor of $\sim 2$ larger $L_{\rm X}^{\prime}$(SW) of $3.9\times 10^{42}$ \ergps. This is a special configuration and it sets the upper bound of $L_{\rm X}^{\prime}$ for {\it SW-A}.

% Fig Lx SW curve vs inclination

\begin{figure}
\centerline{\includegraphics[width=0.4\textwidth,angle=0]{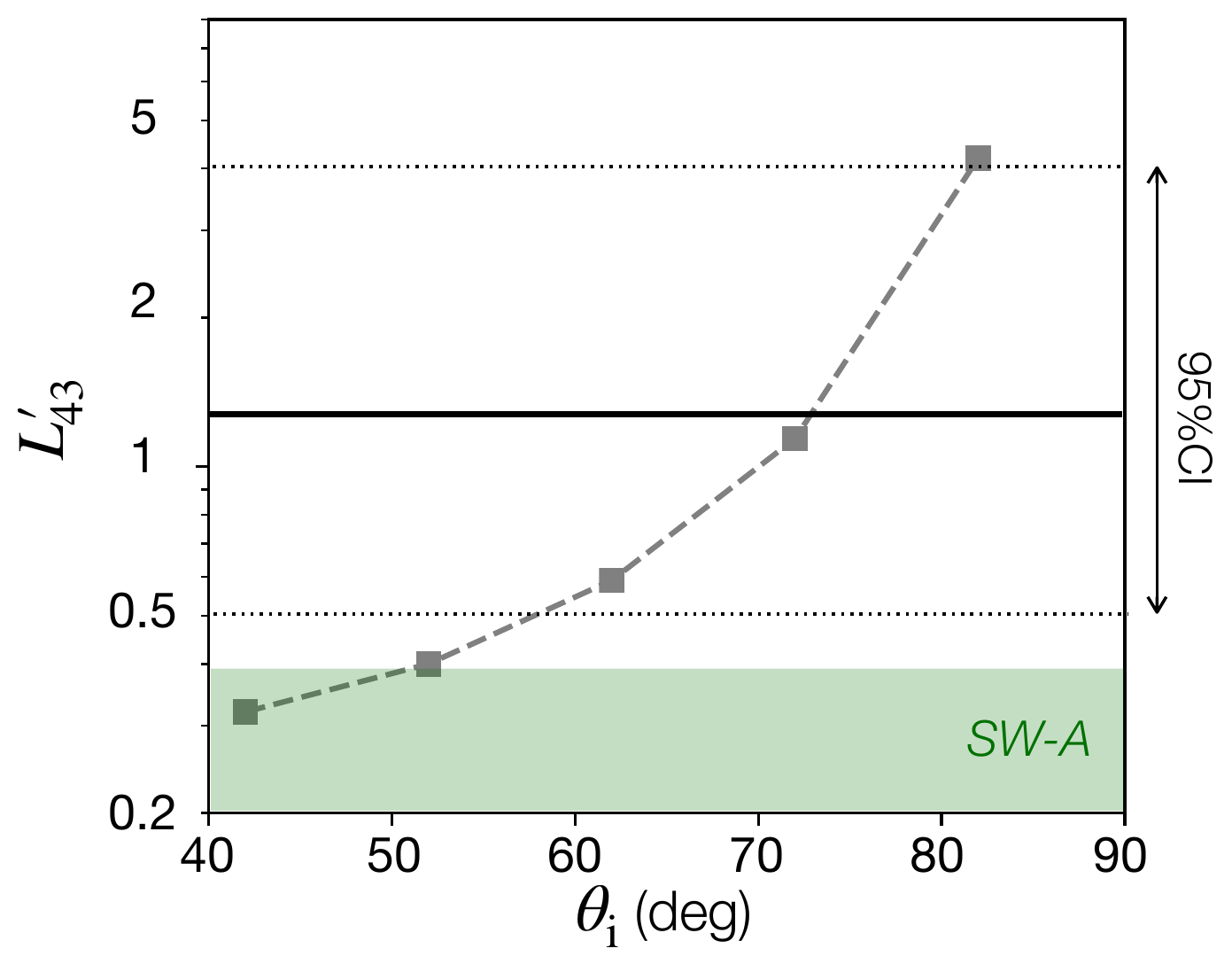}}
\caption{Intrinsic 2-10 keV luminosity of the SW nucleus estimated for the reflection-dominated model {\it SW-R}, as a function of assumed inclination angle ($\theta_\mathrm{i}$) of the obscuring torus. The intrinsic luminosity, $L^{\prime}_{43}$, is in units of $10^{43}$ \ergps. For $\theta_\mathrm{i}=42^{\circ}$-72$^{\circ}$, $\theta_\mathrm{i}=\theta_\mathrm{op}+2^{\circ}$ is assumed. For the point for $\theta_\mathrm{i}=82^{\circ}$, $\theta_\mathrm{op}=70^{\circ}$ is used as it is the upper bound of the {\tt etorus} model. The thick horizontal line indicates the best-estimate of $L_{\rm X}^{\prime} = 1.3\times 10^{43}$ \ergps, which was obtained by combining the X-ray and mid-IR feature based estimates (see Sect. 4.1) and the 95\% CI is indicated by two dotted-lines. The $L_{\rm X}^{\prime}$ value derived from the nominal {\it SW-A} model (see Table \ref{tab:nu_fit}) is $0.2\times 10^{43}$ \ergps, i.e. the bottom of the panel, but it could go up within the shaded area depending on an assumed geometry (see text) in the absorption model.}
\label{fig:lxswcurve}
\end{figure}

Uncertainty in the intrinsic luminosity estimate is even larger when observed light is reflection-dominated as no direct light from the obscured source is visible. In the {\it SW-R} model, $\theta_\mathrm{op}=70^{\circ}$ and $\theta_\mathrm{i}=72^{\circ}$ are assumed, but as long as $\theta_\mathrm{i} = \theta_\mathrm{op}+2^{\circ}$ is maintained, $\theta_\mathrm{op} = 40^{\circ}$-$70^{\circ}$ gives a similar quality of fit. However, since the visible surface of the torus inner wall becomes small as the torus inclination $\theta_\mathrm{i}$ increases (while the spectral shape of reflected light remains similar), the illuminating source needs to be more luminous to produce the same reflection luminosity and vice versa, as is shown in Fig. \ref{fig:lxswcurve}. For $\theta_\mathrm{i} = 72^{\circ}$, the instrinsic luminosity was estimated to be $L_{\rm X}^{\prime} = 1.1\times 10^{43}$ \ergps, but it could be smaller if $\theta_\mathrm{i}$ were smaller. At a given $\theta_\mathrm{i}$, decreasing $\theta_\mathrm{op}$ also makes the visible reflection surface smaller and thus pushes the illuminating luminosity up. However, if $\theta_\mathrm{op}$ is too small, it would loose the quality of fit for the NuSTAR data due to a mismatch in the Fe K absorption edge depth; a small $\theta_\mathrm{op}$ results in a relative increase in reflection light coming through the torus %(``Reflection 1'' in \citet{ikeda09})
over directly visible reflection
%(``Reflection 2'')
leading to a deeper Fe K edge, which overestimates the observed edge depth. For $\theta_\mathrm{i}=72^{\circ}$, we obtained a constraint of $\theta_\mathrm{op}> 55^{\circ}$ (the 90\% lower limit). By decreasing $\theta_\mathrm{op}$, the data start to favour the absorption-dominated spectrum with a smaller \nH, as is similar to what was obtained in {\it SW-A}.
Thus, although there is a small forbidden area in the parameter space, various combinations of the torus parameters are possible to describe the NuSTAR spectrum.
Accordingly, the 2-10 keV intrinsic luminosity of the obscured nucleus in SW, $L_{\rm X}^{\prime}$(SW), could range from $2\times 10^{42}$ \ergps\ to $4\times 10^{43}$ \ergps.

\section{Discussion}

% Table  -- AGN fraction

\begin{table}
\caption{AGN luminosities of the NE and SW nuclei.}
\label{tab:fagn}
\centering
\begin{tabular}{ccccc}
  \hline\hline
 & log $L_{\rm X}^{\prime} $ & log $L_{\rm AGN}$ & log $\lambda_{\rm Edd}$ & log ($L_{\rm AGN}/L_{\rm IR}$) \\
  \hline
%NE & 6e42 & 2e-4 & 2.2e-2 \\
%SW & 1.6e44 (3e43-6e45) & 5e-2 (1e-3-2e-1) & 1.8e-1 (3.4e-2 - 0.69)\\
  NE & 41.60 & 42.80 & $-3.68$ & $-1.63$ \\
  SW & 43.13 & 44.43 & $-2.04$ & $-0.51$ \\
   \hline
\end{tabular}
\tablefoot{The best-estimates for the intrinsic 2-10 keV luminosity $L_{\rm X}^{\prime}$  and X-ray based AGN bolometric luminosity $L_{\rm AGN}$ in units of \ergps, Eddington ratio $\lambda_{\rm Edd}$, and the AGN luminosity and IR luminosity ratio for the NE and SW nuclei. See text and Fig. \ref{fig:lxswcurve} for details on the estimates of $L_{\rm X}^{\prime}$, their uncertainties, and the bolometric corrections.}
\end{table}

Our analysis of the NuSTAR spectrum suggests that the SW nucleus contains a Compton-thick nucleus, which is intrinsically more luminous than the NE nucleus, as speculated by \citet{mazzarella12}. Before examining their AGN properties in turn, some basic parameters of the two nuclei are summarised. The black hole masses for the two nuclei, which were estimated from their host spheroid luminosities, are similarly $M_{\rm BH}=2\times 10^8 M_{\odot}$, for which the Eddington luminosity is $L_{\rm Edd}\sim 3\times 10^{46}$ \ergps. The IR luminosity estimated for the NE and SW nuclei are $L_{\rm IR}{\rm (NE)}=2.7\times 10^{44}$ \ergps\  and $L_{\rm IR}{\rm (SW)}=8.7\times 10^{44}$ \ergps. The AGN bolometric luminosity estimated from X-ray $L_\mathrm{AGN} = \kappa L_{\rm X}^{\prime}$ where $\kappa$ is the X-ray bolometric correction, along with the Eddington ratio $\lambda_\mathrm{Edd}$, and the AGN and IR luminosity ratio $L_\mathrm{AGN}/L_\mathrm{IR}$ for each nucleus, as discussed below, are given in Table \ref{tab:fagn}.

\subsection{SW nucleus}

As shown in Sect. 3, the SW nucleus is likely a Compton-thick AGN with \nH $\geq 2\times 10^{24}$ \psqcm, but the intrinsic X-ray luminosity $L_{\rm X}^{\prime}$ of the heavily obscured source remains uncertain between $ 2\times 10^{42}$ \ergps\ and $4\times 10^{43}$ \ergps.
AGN-related mid-IR diagnostics seem to have a close relation to AGN bolometric power \citep[e.g.][]{gruppioni2016}. Since some of them have been reported to have a good correlation directly with intrinsic X-ray emission, we examine their predictions for$L_{\rm X}^{\prime}$ below.

On account of being dust re-radiation from an obscuring torus, the mid-infrared continuum emission of AGN has been found to tightly correlate with the intrinsic 2-10 keV luminosity \citep{lutz2004,horst2008,gandhi2009,mateos2015}. However, one difficulty is isolating an AGN component from other continuum sources at the same wavelength. In Mrk 266, the available mid-IR photometric data were obtained from apertures with greater than a few arcsecs of the Spitzer-IRAC or IRAS, and we faced the difficulty mentioned above. First, we directly adopted the available photometric data, which gives an upper limit on $L_{\rm X}^{\prime}$. From the SED given in Fig. 9 of \citet{mazzarella12}, luminosities of the SW nucleus at 6 $\mu $m and 12 $\mu$m were estimated to be $L_6 = 4.2\times 10^{43}$ \ergps\ and $L_{12} = 1.0\times 10^{44}$ \ergps, respectively. Using the correlation for 6 $\mu$m by \citet{mateos2015}, $L_{\rm X}^{\prime}\simeq 2\times 10^{43}f_{\rm AGN}$ \ergps \ was obtained, where $f_{\rm AGN}$ denotes the AGN fraction at the mid-IR wavelength. Similarly, the correlation for 12 $\mu$m of \citet{gandhi2009} gave $L_{\rm X}^{\prime}\simeq 5\times 10^{43} f_{\rm AGN}$ \ergps. We note that \citet{mateos2015} and \citet{gandhi2009} used distinct samples of AGN and their scaling-relations were calibrated independently. The AGN fraction, $f_{\rm AGN}$, in the SW nucleus was estimated using various mid-IR diagnostics \citep{mazzarella12,diaz-santos2017} but the obtained values of $f_{\rm AGN}$ spread widely, likely due to aperture effects. Here, we tentatively adopt the median of the six diagnostics, $f_{\rm AGN}=0.35$, which yields $L_{\rm X}^{\prime}$ estimates of $0.7\times 10^{43}$ \ergps\ and $1.7\times 10^{43}$ \ergps, based on $L_6$ and $L_{12}$, respectively.

Another X-ray luminosity estimator is the [O {\sc iv}] $\lambda 25.9\mu $m emission line, which also shows a tight correlation with $L_{\rm X}$ \citep{melendez2008,rigby2009,diamond-stanic2009,liu2014}. As [O {\sc iv}] is a high-excitation line, which likely arises from AGN irradiation, it should be largely free from external contamination \citep{pereira-santaella2010}, unlike the continuum emission; therefore, it directly scales with the AGN luminosity. Being a mid-IR line, it is also relatively robust against dust extinction. Therefore it is deemed to give a reliable estimate of the AGN intrinsic luminosity. The [O {\sc iv}] flux measured with the IRS aperture only for the SW nucleus is available \citep{mazzarella12,bernard-salas2009}, which gives [O {\sc iv}] luminosity of $8\times 10^{41}$ \ergps. The [O{\sc iv}]-$L_{\rm X}^{\prime}$ correlation of \citet{liu2014} for Seyfert 2 galaxies gives $L_{\rm X}^{\prime}\simeq 1.9\times 10^{43}$ \ergps.

The above three predictions for $L_{\rm X}^{\prime}$ from the mid-IR features all lie around $\sim 1\times 10^{43}$ \ergps\  and within the range inferred by the modelling of the NuSTAR data. This supports the reflection-dominated spectrum of {\it SW-R} for modelling the data (see Fig. \ref{fig:lxswcurve}). By combining these estimates, we looked into the most likely value of $L_{\rm X}^{\prime}$(SW) and its uncertainty and then assessed how probable the reflection-dominated hypothesis might be. We label the X-ray and the mid-IR estimates as E$i$ ($i=0,1,2,3$), where E0 is NuSTAR spectral modelling, E1 is $6 \mu$m based, E2 is $12 \mu$m based, and E3 is [O {\sc iv}] based. In E0, we assume that log$L_{\rm X}^{\prime}$ is equally likely between 42.3 and 43.6 [\ergps], or E0$\sim Unif(42.3, 43.6)$. The mid-IR features-based estimates have uncertainties due to scatter around each correlation: 0.4 dex for E1 and 0.23 dex for E2 are given by \citet{mateos2015} and \citet{gandhi2009}, respectively, and we estimate 0.5 dex for E3. We assume that each estimate of log $L_{\rm X}^{\prime} (\equiv\mu_i)$ follows a Gaussian distribution of E$i\sim N(\mu_i, \sigma_i)$ ($i=1,2,3$): E1$\sim N(42.85, 0.5)$, E2$\sim N(43.17, 0.38)$, and E3$\sim N(43.19, 0.5)$, for which we included a factor of two of uncertainty of $f_{\rm AGN}$ to $\sigma_1$ and $\sigma_2$ in addition to the scatter of the correlations. The posterior probability $\prod {\rm E}i$ gives the most likely estimate of log $L_{\rm X}^{\prime} =  43.13$ (the 95\% compatible interval (CI): 42.7-43.6). This is illustrated in Fig. \ref{fig:lxswcurve}. The $L_{\rm X}^{\prime}$ range that can be obtained in {\it SW-A} lies below the 95\% CI, indicating that {\it SW-R} is preferred and a reflection-dominated spectrum is a more probable description of the NuSTAR data for the SW nucleus.

Assuming a bolometric correction of 20 \citep{marconi04}, the $L_{\rm AGN}$ (SW) is found to be $2.6\times 10^{44}$ \ergps, based on the best-estimate of $L_{\rm X}^{\prime}$. The Eddington ratio is then 0.9\%, which is typical of Seyferts. The luminosity ratio $L_{\rm AGN}/L_{\rm IR}\simeq 0.3$ can be translated to the AGN fraction in the galaxy's bolometric output if $L_{\rm AGN}$ is all absorbed by circumnuclear dust and reradiated in the IR.

\subsection{NE nucleus}

The X-ray luminosity of the NE nucleus is relatively low, about $4\times 10^{41}$ \ergps. Although the hard Chandra spectrum indicates a significant AGN contribution in the Chandra band \citep{torres-alba2018} since intense star formation is also taking place in the NE nucleus, we first examine how much the star-forming activity might contribute to the observed hard X-ray emission. At energies above 3 keV, starburst emission mainly comes from high-mass X-ray binaries. Their collective X-ray luminosity can be estimated using the empirical relation with the star formation rate \citep{grimm2003,ranalli2003,lehmer2010}. The star formation rate of the entire Mrk 266 system is 65 \Msun\ yr$^{-1}$ \citep{howell2010}. By scaling with $L_{\rm IR}$, we estimate the star formation rate of the NE nucleus to be 15 \Msun\ yr$^{-1}$. Using the formula of \citet{lehmer2010}  and given the star formation rate, $L_{\rm X}$, is dominated by high-mass X-ray binaries and given the contribution of low-mass X-ray binaries is minor, the expected 2-10 keV luminosity from X-ray binaries is found to be $2.4\times 10^{40}$ \ergps\  or $\sim 5$ \% of the observed 2-10 keV luminosity. The radial surface brightness profile of the NE nucleus in the 4-7 keV Chandra image (Fig. \ref{fig:hstimg}) shows a small extension ($\leq 10$\% of the total emission), which might be attributed to the star formation, but most of the emission comes from a low-luminosity AGN. The 3-7 keV flux observed by Chandra is $\sim $ 25\% higher than that of XMM-Newton and NuSTAR. Although the increase is marginal, X-ray variability would support the dominance of AGN emission. 
%(Fig. \ref{fig:nesurbri})
 
The Eddington ratio of NE is rather low ($\lambda_{\rm Edd}\sim 2\times 10^{-4}$). Among nearby AGN, the $\lambda_{\rm Edd}$ range around this value is populated by LINERs and low-luminosity Seyferts \citep{ho2008araa}. Although the radio to mid-IR flux-density ratio, $q24obs = {\rm log}(S_{\rm 1.4GHz}/S_{24 \mu {\rm m}}) = 1.1$, is consistent with those of star forming galaxies \citep{sargent2010}, the radio source is compact and unresolved at HPBW $= 0.^{\prime\prime}3\times 0.^{\prime\prime}4$ (\citealp{mazzarella1988}). Furthermore, it is likely associated with the AGN, which may have an inefficient accretion flow, suggested by the low $\lambda_{\rm Edd}$, producing the radio emission. The observed radio flux of the NE nucleus is comparable with that of the SW nucleus with similar spectral slopes: $-0.9$ for the NE nucleus and $-0.8$ for the SW nucleus in the 1.4-15 GHz range \citep{mazzarella1988}, whereas the intrinsic $L_{\rm X}$ is $\sim 1.6$ dex lower than the SW nucleus (Table \ref{tab:fagn}). This seems to be compatible with the characteritic radio excess of low-luminosity AGN (LLAGN) with an inefficient accretion flow \citep{ho2008araa}.

When the bolometric correction $\kappa = 16$, suggested for LLAGN by \citet{ho2008araa} is used, $L_{\rm AGN} = 6\times 10^{42}$ \ergps\ is obtained. The $L_{\rm AGN}/L_{\rm IR}$ is therefore $\sim 2$ \%. The AGN contribution to the IR luminosity could even be lower if the NE nucleus had a SED typical of LLAGN, lacking a UV excess, which is a main source of heating of IR emitting circumnuclear dusts. This is in stark contrast to $f_{\rm AGN}\sim 60$\% obtained from the mid-IR diagnostics \citep{mazzarella12}. However, the Spitzer IRS spectrum of the NE nucleus alone was taken only from the SL module and it lacks the longer wavelength coverage where the key AGN diagnostic lines [Ne {\sc v}]$\lambda 14.3, 24.3 \mu$m and [O {\sc iv}]$\lambda 25.9\mu$m are present. This leaves the PAH strengths as the only mid-IR AGN diagnostics that may be biased.  Given the low accretion rate inferred from the X-ray observations, strong outflows are expected from the AGN in the NE nucleus, which fit observed optical and radio signatures pointed out by \citet{mazzarella12}. The AGN energy output of the NE nucleus may be dominated by the mechanical power of those outflowing winds.

\subsection{Dual AGN in Mrk 266}

Mrk 266 exhibits properties that have close relevance to the
merger-driven formation of dual AGN and their detectability.
Two important characteristics of this merger system are
1) a close nuclear separation of 6 kpc in projection; and
2) the mass ratio of the host spheroids are close to unity and thus, presumably, as is the black hole mass ratio.
These two are required conditions for forming dual AGN deduced from cosmological simulations \citep[][see also e.g. \citealp{hopkins2006, solanes2019}]{steinborn2016,volonteri2016}. As a substantial amount of cold molecular gas ($\sim $a few $10^9$ \Msun) is available between the galaxies \citep{imanishi09}, this reservoir provides individual galaxies with gas for further accretion, helping the system evolve into an ULIRG, as argued by \citet{mazzarella12}. However, it is interesting to note that the two AGN appear to have distinct characteristics: The SW nucleus is $\simeq 30$ times more luminous than the NE nucleus and they are possibly accreting in different modes as discussed above, suggesting that the accretion conditions might actually be unbalanced between the galaxies. In terms of detectability of dual AGN, heavy obscuration certainly plays a role. \citet{ricci2017} found an elevated proportion of Compton-thick AGN in advanced mergers with nuclear separations of $<10$ kpc. Mrk 266 fits in the merger stage and the SW nucleus is indeed a Compton-thick AGN, which had not been verified until the NuSTAR observation. The spatial resolution of Chandra and the hard X-ray sensitivity of NuSTAR complement each other, helping to identify a Compton-thick AGN in advanced merger systems such as Mrk 266 and Mrk 273. A multiwavelength approach, as employed by GOALS, is obviously effective at overcoming the obscuration issue \citep[e.g.][]{hickox2018}, which would naturally be expected in luminous merger systems. 
We refer readers to an extensive review of both observational and theoretical aspects of dual AGN by \citet{derosa2019} who further discuss related X-ray observational works on samples that were selected using various techniques \citep[e.g.][]{koss2010,koss2012,comerford2015,satyapal2017}.

\begin{acknowledgements}
This research made use of data obtained from NuSTAR, XMM-Newton and Chandra X-ray Observatory, software packages of HEASoft and R \citep{rcoreteam2017}, and the NASA/IPAC Extragalactic Databases (NED), which is funded by the National Aeronatutics and Space Administration and operated by the California Institute of Technology. KI acknowledges support by the Spanish MICINN under grant PID2019-105510GB-C33. T.D-S. acknowledges support from the CASSACA and CONICYT fund CAS-CONICYT Call 2018. C.R. acknowledges support from the Fondecyt Iniciacion grant 11190831. ASE, GCP and KI acknowledge NASA Astrophysics Data Analysis Program (ADAP) Grant 80NSSC20K0450 (PI: U).
\end{acknowledgements}

\bibliographystyle{aa} \bibliography{mrk266}{}

\begin{thebibliography}{59}
\expandafter\ifx\csname natexlab\endcsname\relax\def\natexlab#1{#1}\fi

\bibitem[{{Armus} {et~al.}(2009){Armus}, {Mazzarella}, {Evans}, {Surace},
  {Sanders}, {Iwasawa}, {Frayer}, {Howell}, {Chan}, {Petric}, {Vavilkin},
  {Kim}, {Haan}, {Inami}, {Murphy}, {Appleton}, {Barnes}, {Bothun}, {Bridge},
  {Charmandaris}, {Jensen}, {Kewley}, {Lord}, {Madore}, {Marshall},
  {Melbourne}, {Rich}, {Satyapal}, {Schulz}, {Spoon}, {Sturm}, {U}, {Veilleux},
  \& {Xu}}]{armus09}
{Armus}, L., {Mazzarella}, J.~M., {Evans}, A.~S., {et~al.} 2009, \pasp, 121,
  559

\bibitem[{{Awaki} {et~al.}(2009){Awaki}, {Terashima}, {Higaki}, \&
  {Fukazawa}}]{awaki09}
{Awaki}, H., {Terashima}, Y., {Higaki}, Y., \& {Fukazawa}, Y. 2009, \pasj, 61,
  S317

\bibitem[{{Balokovi{\'c}} {et~al.}(2018){Balokovi{\'c}}, {Brightman},
  {Harrison}, {Comastri}, {Ricci}, {Buchner}, {Gandhi}, {Farrah}, \&
  {Stern}}]{balokovic2018}
{Balokovi{\'c}}, M., {Brightman}, M., {Harrison}, F.~A., {et~al.} 2018, \apj,
  854, 42

\bibitem[{{Bernard-Salas} {et~al.}(2009){Bernard-Salas}, {Spoon},
  {Charmandaris}, {Lebouteiller}, {Farrah}, {Devost}, {Brand l}, {Wu}, {Armus},
  {Hao}, {Sloan}, {Weedman}, \& {Houck}}]{bernard-salas2009}
{Bernard-Salas}, J., {Spoon}, H.~W.~W., {Charmandaris}, V., {et~al.} 2009,
  \apjs, 184, 230

\bibitem[{{Brassington} {et~al.}(2007){Brassington}, {Ponman}, \&
  {Read}}]{brassington2007}
{Brassington}, N.~J., {Ponman}, T.~J., \& {Read}, A.~M. 2007, \mnras, 377, 1439

\bibitem[{{Chu} {et~al.}(2017){Chu}, {Sanders}, {Larson}, {Mazzarella},
  {Howell}, {D{\'\i}az-Santos}, {Xu}, {Paladini}, {Schulz}, {Shupe},
  {Appleton}, {Armus}, {Billot}, {Chan}, {Evans}, {Fadda}, {Frayer}, {Haan},
  {Ishida}, {Iwasawa}, {Kim}, {Lord}, {Murphy}, {Petric}, {Privon}, {Surace},
  \& {Treister}}]{chu2017}
{Chu}, J.~K., {Sanders}, D.~B., {Larson}, K.~L., {et~al.} 2017, \apjs, 229, 25

\bibitem[{{Comerford} {et~al.}(2015){Comerford}, {Pooley}, {Barrows}, {Greene},
  {Zakamska}, {Madejski}, \& {Cooper}}]{comerford2015}
{Comerford}, J.~M., {Pooley}, D., {Barrows}, R.~S., {et~al.} 2015, \apj, 806,
  219

\bibitem[{{De Rosa} {et~al.}(2019){De Rosa}, {Vignali}, {Bogdanovi{\'c}},
  {Capelo}, {Charisi}, {Dotti}, {Husemann}, {Lusso}, {Mayer}, {Paragi},
  {Runnoe}, {Sesana}, {Steinborn}, {Bianchi}, {Colpi}, {del Valle}, {Frey},
  {Gab{\'a}nyi}, {Giustini}, {Guainazzi}, {Haiman}, {Herrera Ruiz},
  {Herrero-Illana}, {Iwasawa}, {Komossa}, {Lena}, {Loiseau}, {Perez-Torres},
  {Piconcelli}, \& {Volonteri}}]{derosa2019}
{De Rosa}, A., {Vignali}, C., {Bogdanovi{\'c}}, T., {et~al.} 2019, \nar, 86,
  101525

\bibitem[{{Diamond-Stanic} {et~al.}(2009){Diamond-Stanic}, {Rieke}, \&
  {Rigby}}]{diamond-stanic2009}
{Diamond-Stanic}, A.~M., {Rieke}, G.~H., \& {Rigby}, J.~R. 2009, \apj, 698, 623

\bibitem[{{D{\'\i}az-Santos} {et~al.}(2017){D{\'\i}az-Santos}, {Armus},
  {Charmandaris}, {Lu}, {Stierwalt}, {Stacey}, {Malhotra}, {van der Werf},
  {Howell}, {Privon}, {Mazzarella}, {Goldsmith}, {Murphy}, {Barcos-Mu{\~n}oz},
  {Linden}, {Inami}, {Larson}, {Evans}, {Appleton}, {Iwasawa}, {Lord},
  {Sanders}, \& {Surace}}]{diaz-santos2017}
{D{\'\i}az-Santos}, T., {Armus}, L., {Charmandaris}, V., {et~al.} 2017, \apj,
  846, 32

\bibitem[{{Gandhi} {et~al.}(2009){Gandhi}, {Horst}, {Smette}, {H{\"o}nig},
  {Comastri}, {Gilli}, {Vignali}, \& {Duschl}}]{gandhi2009}
{Gandhi}, P., {Horst}, H., {Smette}, A., {et~al.} 2009, \aap, 502, 457

\bibitem[{{Grimm} {et~al.}(2003){Grimm}, {Gilfanov}, \& {Sunyaev}}]{grimm2003}
{Grimm}, H.~J., {Gilfanov}, M., \& {Sunyaev}, R. 2003, \mnras, 339, 793

\bibitem[{{Gruppioni} {et~al.}(2016){Gruppioni}, {Berta}, {Spinoglio},
  {Pereira-Santaella}, {Pozzi}, {Andreani}, {Bonato}, {De Zotti}, {Malkan},
  {Negrello}, {Vallini}, \& {Vignali}}]{gruppioni2016}
{Gruppioni}, C., {Berta}, S., {Spinoglio}, L., {et~al.} 2016, \mnras, 458, 4297

\bibitem[{{Hickox} \& {Alexander}(2018)}]{hickox2018}
{Hickox}, R.~C. \& {Alexander}, D.~M. 2018, \araa, 56, 625

\bibitem[{{Ho}(2008)}]{ho2008araa}
{Ho}, L.~C. 2008, \araa, 46, 475

\bibitem[{{Hopkins} {et~al.}(2006){Hopkins}, {Hernquist}, {Cox}, {Di Matteo},
  {Robertson}, \& {Springel}}]{hopkins2006}
{Hopkins}, P.~F., {Hernquist}, L., {Cox}, T.~J., {et~al.} 2006, \apjs, 163, 1

\bibitem[{{Horst} {et~al.}(2008){Horst}, {Gandhi}, {Smette}, \&
  {Duschl}}]{horst2008}
{Horst}, H., {Gandhi}, P., {Smette}, A., \& {Duschl}, W.~J. 2008, \aap, 479,
  389

\bibitem[{{Howell} {et~al.}(2010){Howell}, {Armus}, {Mazzarella}, {Evans},
  {Surace}, {Sanders}, {Petric}, {Appleton}, {Bothun}, {Bridge}, {Chan},
  {Charmandaris}, {Frayer}, {Haan}, {Inami}, {Kim}, {Lord}, {Madore},
  {Melbourne}, {Schulz}, {U}, {Vavilkin}, {Veilleux}, \& {Xu}}]{howell2010}
{Howell}, J.~H., {Armus}, L., {Mazzarella}, J.~M., {et~al.} 2010, \apj, 715,
  572

\bibitem[{{Hutchings} {et~al.}(1988){Hutchings}, {Neff}, \& {van
  Gorkom}}]{hutchings1988}
{Hutchings}, J.~B., {Neff}, S.~G., \& {van Gorkom}, J.~H. 1988, \aj, 96, 1227

\bibitem[{{Ikeda} {et~al.}(2009){Ikeda}, {Awaki}, \& {Terashima}}]{ikeda09}
{Ikeda}, S., {Awaki}, H., \& {Terashima}, Y. 2009, \apj, 692, 608

\bibitem[{{Imanishi} {et~al.}(2009){Imanishi}, {Nakanishi}, {Tamura}, \&
  {Peng}}]{imanishi09}
{Imanishi}, M., {Nakanishi}, K., {Tamura}, Y., \& {Peng}, C.-H. 2009, \aj, 137,
  3581

\bibitem[{{Ishigaki} {et~al.}(2000){Ishigaki}, {Yoshida}, {Aoki}, {Ohtani},
  {Sugai}, {Hayashi}, {Ozaki}, {Hattori}, \& {Ishii}}]{ishigaki2000}
{Ishigaki}, T., {Yoshida}, M., {Aoki}, K., {et~al.} 2000, \pasj, 52, 185

\bibitem[{{Iwasawa} {et~al.}(2011){Iwasawa}, {Mazzarella}, {Surace}, {Sand
  ers}, {Armus}, {Evans}, {Howell}, {Komossa}, {Petric}, {Teng}, {U}, \&
  {Veilleux}}]{iwasawa11}
{Iwasawa}, K., {Mazzarella}, J.~M., {Surace}, J.~A., {et~al.} 2011, \aap, 528,
  A137

\bibitem[{{Iwasawa} {et~al.}(2018){Iwasawa}, {U}, {Mazzarella}, {Medling},
  {Sanders}, \& {Evans}}]{iwasawa18}
{Iwasawa}, K., {U}, V., {Mazzarella}, J.~M., {et~al.} 2018, \aap, 611, A71

\bibitem[{{Kollatschny} \& {Fricke}(1984)}]{kollatschny1984}
{Kollatschny}, W. \& {Fricke}, K.~J. 1984, \aap, 135, 171

\bibitem[{{Koss} {et~al.}(2012){Koss}, {Mushotzky}, {Treister}, {Veilleux},
  {Vasudevan}, \& {Trippe}}]{koss2012}
{Koss}, M., {Mushotzky}, R., {Treister}, E., {et~al.} 2012, \apjl, 746, L22

\bibitem[{{Koss} {et~al.}(2010){Koss}, {Mushotzky}, {Veilleux}, \&
  {Winter}}]{koss2010}
{Koss}, M., {Mushotzky}, R., {Veilleux}, S., \& {Winter}, L. 2010, \apjl, 716,
  L125

\bibitem[{{Lehmer} {et~al.}(2010){Lehmer}, {Alexander}, {Bauer}, {Brand t},
  {Goulding}, {Jenkins}, {Ptak}, \& {Roberts}}]{lehmer2010}
{Lehmer}, B.~D., {Alexander}, D.~M., {Bauer}, F.~E., {et~al.} 2010, \apj, 724,
  559

\bibitem[{{Liu} {et~al.}(2014){Liu}, {Wang}, {Yang}, {Zhu}, \&
  {Zhou}}]{liu2014}
{Liu}, T., {Wang}, J.-X., {Yang}, H., {Zhu}, F.-F., \& {Zhou}, Y.-Y. 2014,
  \apj, 783, 106

\bibitem[{{Liu} {et~al.}(2019){Liu}, {Veilleux}, {Iwasawa}, {Rupke}, {Teng},
  {U}, {Tombesi}, {Sanders}, {Max}, \& {Mel{\'e}ndez}}]{liu2019}
{Liu}, W., {Veilleux}, S., {Iwasawa}, K., {et~al.} 2019, \apj, 872, 39

\bibitem[{{Lutz} {et~al.}(2004){Lutz}, {Maiolino}, {Spoon}, \&
  {Moorwood}}]{lutz2004}
{Lutz}, D., {Maiolino}, R., {Spoon}, H.~W.~W., \& {Moorwood}, A.~F.~M. 2004,
  \aap, 418, 465

\bibitem[{{Madsen} {et~al.}(2020){Madsen}, {Grefenstette}, {Pike}, {Miyasaka},
  {Brightman}, {Forster}, \& {Harrison}}]{madsen2020nustar}
{Madsen}, K.~K., {Grefenstette}, B.~W., {Pike}, S., {et~al.} 2020, arXiv
  e-prints, arXiv:2005.00569

\bibitem[{{Marconi} \& {Hunt}(2003)}]{marconihunt03}
{Marconi}, A. \& {Hunt}, L.~K. 2003, \apjl, 589, L21

\bibitem[{{Marconi} {et~al.}(2004){Marconi}, {Risaliti}, {Gilli}, {Hunt},
  {Maiolino}, \& {Salvati}}]{marconi04}
{Marconi}, A., {Risaliti}, G., {Gilli}, R., {et~al.} 2004, \mnras, 351, 169

\bibitem[{{Mateos} {et~al.}(2015){Mateos}, {Carrera}, {Alonso-Herrero},
  {Rovilos}, {Hern{\'a}n-Caballero}, {Barcons}, {Blain}, {Caccianiga}, {Della
  Ceca}, \& {Severgnini}}]{mateos2015}
{Mateos}, S., {Carrera}, F.~J., {Alonso-Herrero}, A., {et~al.} 2015, \mnras,
  449, 1422

\bibitem[{{Matt} {et~al.}(1999){Matt}, {Pompilio}, \& {La Franca}}]{matt1999}
{Matt}, G., {Pompilio}, F., \& {La Franca}, F. 1999, \na, 4, 191

\bibitem[{{Mazzarella} \& {Boroson}(1993)}]{mazzarellaboroson1993}
{Mazzarella}, J.~M. \& {Boroson}, T.~A. 1993, \apjs, 85, 27

\bibitem[{{Mazzarella} {et~al.}(1988){Mazzarella}, {Gaume}, {Aller}, \&
  {Hughes}}]{mazzarella1988}
{Mazzarella}, J.~M., {Gaume}, R.~A., {Aller}, H.~D., \& {Hughes}, P.~A. 1988,
  \apj, 333, 168

\bibitem[{{MIV12}{}(){Mazzarella}, {Iwasawa}, {Vavilkin},
  {Armus}, {Kim}, {Bothun}, {Evans}, {Spoon}, {Haan}, {Howell}, {Lord},
  {Marshall}, {Ishida}, {Xu}, {Petric}, {Sanders}, {Surace}, {Appleton},
  {Chan}, {Frayer}, {Inami}, {Khachikian}, {Madore}, {Privon}, {Sturm}, {U}, \&
  {Veilleux}}]{mazzarella12}
{Mazzarella}, J.~M., {Iwasawa}, K., {Vavilkin}, T., {et~al.} 2012, \aj, 144,
  125

\bibitem[{{Mel{\'e}ndez} {et~al.}(2008){Mel{\'e}ndez}, {Kraemer}, {Armentrout},
  {Deo}, {Crenshaw}, {Schmitt}, {Mushotzky}, {Tueller}, {Markwardt}, \&
  {Winter}}]{melendez2008}
{Mel{\'e}ndez}, M., {Kraemer}, S.~B., {Armentrout}, B.~K., {et~al.} 2008, \apj,
  682, 94

\bibitem[{{Nandra} \& {Pounds}(1994)}]{nandrapounds94}
{Nandra}, K. \& {Pounds}, K.~A. 1994, \mnras, 268, 405

\bibitem[{{Osterbrock} \& {Dahari}(1983)}]{osterbrock1983}
{Osterbrock}, D.~E. \& {Dahari}, O. 1983, \apj, 273, 478

\bibitem[{{Osterbrock} \& {Martel}(1993)}]{osterbrockmartel1993}
{Osterbrock}, D.~E. \& {Martel}, A. 1993, \apj, 414, 552

\bibitem[{{Pereira-Santaella} {et~al.}(2010){Pereira-Santaella},
  {Diamond-Stanic}, {Alonso-Herrero}, \& {Rieke}}]{pereira-santaella2010}
{Pereira-Santaella}, M., {Diamond-Stanic}, A.~M., {Alonso-Herrero}, A., \&
  {Rieke}, G.~H. 2010, \apj, 725, 2270

\bibitem[{{R Core Team}(2017)}]{rcoreteam2017}
{R Core Team}. 2017, R: A Language and Environment for Statistical Computing, R
  Foundation for Statistical Computing, Vienna, Austria

\bibitem[{{Ranalli} {et~al.}(2003){Ranalli}, {Comastri}, \&
  {Setti}}]{ranalli2003}
{Ranalli}, P., {Comastri}, A., \& {Setti}, G. 2003, \aap, 399, 39

\bibitem[{{Ricci} {et~al.}(2017){Ricci}, {Bauer}, {Treister}, {Schawinski},
  {Privon}, {Blecha}, {Arevalo}, {Armus}, {Harrison}, {Ho}, {Iwasawa},
  {Sanders}, \& {Stern}}]{ricci2017}
{Ricci}, C., {Bauer}, F.~E., {Treister}, E., {et~al.} 2017, \mnras, 468, 1273

\bibitem[{{Rigby} {et~al.}(2009){Rigby}, {Diamond-Stanic}, \&
  {Aniano}}]{rigby2009}
{Rigby}, J.~R., {Diamond-Stanic}, A.~M., \& {Aniano}, G. 2009, \apj, 700, 1878

\bibitem[{{Sanders} {et~al.}(1986){Sanders}, {Scoville}, {Young}, {Soifer},
  {Schloerb}, {Rice}, \& {Danielson}}]{sanders1986}
{Sanders}, D.~B., {Scoville}, N.~Z., {Young}, J.~S., {et~al.} 1986, \apjl, 305,
  L45

\bibitem[{{Sargent} {et~al.}(2010){Sargent}, {Schinnerer}, {Murphy}, {Aussel},
  {Le Floc'h}, {Frayer}, {Mart{\'\i}nez-Sansigre}, {Oesch}, {Salvato},
  {Smol{\v{c}}i{\'c}}, {Zamorani}, {Brusa}, {Cappelluti}, {Carilli}, {Carollo},
  {Ilbert}, {Kartaltepe}, {Koekemoer}, {Lilly}, {Sand ers}, \&
  {Scoville}}]{sargent2010}
{Sargent}, M.~T., {Schinnerer}, E., {Murphy}, E., {et~al.} 2010, \apjs, 186,
  341

\bibitem[{{Satyapal} {et~al.}(2017){Satyapal}, {Secrest}, {Ricci}, {Ellison},
  {Rothberg}, {Blecha}, {Constantin}, {Gliozzi}, {McNulty}, \&
  {Ferguson}}]{satyapal2017}
{Satyapal}, S., {Secrest}, N.~J., {Ricci}, C., {et~al.} 2017, \apj, 848, 126

\bibitem[{{Solanes} {et~al.}(2019){Solanes}, {Perea}, {Valent{\'\i}-Rojas},
  {del Olmo}, {M{\'a}rquez}, {Ramos Almeida}, \& {Tous}}]{solanes2019}
{Solanes}, J.~M., {Perea}, J.~D., {Valent{\'\i}-Rojas}, G., {et~al.} 2019,
  \aap, 624, A86

\bibitem[{{Steinborn} {et~al.}(2016){Steinborn}, {Dolag}, {Comerford},
  {Hirschmann}, {Remus}, \& {Teklu}}]{steinborn2016}
{Steinborn}, L.~K., {Dolag}, K., {Comerford}, J.~M., {et~al.} 2016, \mnras,
  458, 1013

\bibitem[{{Torres-Alb{\`a}} {et~al.}(2018){Torres-Alb{\`a}}, {Iwasawa},
  {D{\'\i}az-Santos}, {Charmandaris}, {Ricci}, {Chu}, {Sand ers}, {Armus},
  {Barcos-Mu{\~n}oz}, {Evans}, {Howell}, {Inami}, {Linden}, {Medling},
  {Privon}, {U}, \& {Yoon}}]{torres-alba2018}
{Torres-Alb{\`a}}, N., {Iwasawa}, K., {D{\'\i}az-Santos}, T., {et~al.} 2018,
  \aap, 620, A140

\bibitem[{{U} {et~al.}(2013){U}, {Medling}, {Sanders}, {Max}, {Armus},
  {Iwasawa}, {Evans}, {Kewley}, \& {Fazio}}]{u13}
{U}, V., {Medling}, A., {Sanders}, D., {et~al.} 2013, \apj, 775, 115

\bibitem[{{Ueda} {et~al.}(2014){Ueda}, {Akiyama}, {Hasinger}, {Miyaji}, \&
  {Watson}}]{ueda14}
{Ueda}, Y., {Akiyama}, M., {Hasinger}, G., {Miyaji}, T., \& {Watson}, M.~G.
  2014, \apj, 786, 104

\bibitem[{{Volonteri} {et~al.}(2016){Volonteri}, {Dubois}, {Pichon}, \&
  {Devriendt}}]{volonteri2016}
{Volonteri}, M., {Dubois}, Y., {Pichon}, C., \& {Devriendt}, J. 2016, \mnras,
  460, 2979

\bibitem[{{Wang} {et~al.}(1997){Wang}, {Heckman}, {Weaver}, \&
  {Armus}}]{wang1997}
{Wang}, J., {Heckman}, T.~M., {Weaver}, K.~A., \& {Armus}, L. 1997, \apj, 474,
  659

\bibitem[{{Wu} {et~al.}(1998){Wu}, {Zou}, {Xia}, \& {Deng}}]{wu1998}
{Wu}, H., {Zou}, Z.~L., {Xia}, X.~Y., \& {Deng}, Z.~G. 1998, \aaps, 127, 521

\end{thebibliography}

\end{document}